\documentclass[conference]{IEEEtran}
\IEEEoverridecommandlockouts
\usepackage{cite}
\usepackage{amsmath,amssymb,amsfonts}
\usepackage{algorithmic}
\usepackage{graphicx}
\usepackage{textcomp}
\usepackage[table]{xcolor}

\usepackage{enumitem}
\PassOptionsToPackage{hyphens}{url}\usepackage[colorlinks=true,linkcolor=black,anchorcolor=black,citecolor=black,filecolor=black,menucolor=black,runcolor=black,urlcolor=black]{hyperref}
\usepackage[numbers]{natbib}

\def\BibTeX{{\rm B\kern-.05em{\sc i\kern-.025em b}\kern-.08em
    T\kern-.1667em\lower.7ex\hbox{E}\kern-.125emX}}
\begin{document}
\sloppy

\title{Scoping Sustainable Collaborative Mixed Reality}

\author{\IEEEauthorblockN{Yasra Chandio$^{\dagger}$, Noman Bashir$^{\ddagger}$, Tian Guo$^{\intercal}$, Elsa Olivetti$^{\ddagger}$, Fatima M. Anwar$^{\dagger}$}
\IEEEauthorblockA{$^{\dagger}$University of Massachusetts Amherst, $^{\ddagger}$MIT, $^{\intercal}$Worcester Polytechnic Institute}
}

\maketitle

\begin{abstract}
Mixed Reality (MR) is becoming ubiquitous as it finds its applications in education, healthcare, and other sectors beyond leisure. While MR end devices, such as headsets, have low energy intensity, the total number of devices and resource requirements of the entire MR ecosystem, which includes cloud and edge endpoints, can be significant. The resulting operational and embodied carbon footprint of MR has led to concerns about its environmental implications. Recent research has explored reducing the carbon footprint of MR devices by exploring hardware design space or network optimizations. However, many additional avenues for enhancing MR's sustainability remain open, including energy savings in non-processor components and carbon-aware optimizations in collaborative MR ecosystems. In this paper, we aim to identify key challenges, existing solutions, and promising research directions for improving MR sustainability. We explore adjacent fields of embedded and mobile computing systems for insights and outline MR-specific problems requiring new solutions. We identify the challenges that must be tackled to enable researchers, developers, and users to avail themselves of these opportunities in collaborative MR systems.
\end{abstract}

\section{Introduction}
\label{sec:intro}
Mixed Reality (MR) is an emerging technology increasingly used for leisure and safety-critical collaborative applications~\cite{surgery_review, solitary-jogging}. From 2020 to 2021, 33 million AR/VR headsets were sold, a trend poised to accelerate with the release of Apple Vision Pro~\cite{apple-pro-vision-sales} and Generative AI fostering new applications. If used daily for 2 hours, the 33 million headsets could generate 2.6 $\times$10$^5$ metric tons of greenhouse gas (GHG) emissions annually, based on a 50Wh daily usage and the average global carbon intensity of 432\texttt{gCO$_{2}$eq/kWh}~\cite{ember-elec-report}. The carbon footprint of the broader MR ecosystem, including edge computing systems and cloud datacenters, will be much higher than low-power MR headsets.

Recent efforts have explored opportunities to improve the sustainability of MR~\cite{metaverse-networking, Elgamal:2023:Design}. \citet{metaverse-networking} explore optimizing the energy efficiencies of networking components to enable sustainable development in Metaverse. \citet{Elgamal:2023:Design} investigate MR hardware design space optimizations to reduce the lifecycle emissions of a single headset. Prior work on the energy efficiency of MR headsets has explored energy-efficient video processing, optimizing display power, and reducing the power used for tracking, among other applications~\cite{display-power-2022}. There is also work on estimating and optimizing the energy consumption of gaming, which may apply to MR headsets~\cite{mills2019toward}. While prior work takes essential initial steps, significant avenues for improving the sustainability of \emph{collaborative MR} remain open, including carbon-aware spatiotemporal workload optimizations, reducing the carbon footprint of non-processing components, and leveraging prior work in the adjacent fields of embedded and mobile computing systems. 

While prior work can inform sustainability efforts in collaborative MR, reducing MR's energy and carbon footprint involves additional challenges due to the nature of computing tasks in MR. For instance, real-time visual processing for immersive environments is an especially demanding task~\cite{luong2023controllersvshand}. Additionally, continuous user tracking through interaction modalities like hand and eye tracking presents unique challenges not encountered in other domains~\cite{Li:2017:LowPowerGaze}. MR devices' need for portability and wearability brings specific design and operational constraints, such as highly effective thermal management. It may also necessitate task offloading to edge and cloud systems for applications that could otherwise run efficiently on smartphones. These challenges are in addition to the usual issues faced by interactive, battery-powered, and mobile devices, such as balancing battery life with performance and ensuring reliable wireless connectivity.

This paper identifies the potential opportunities for energy- and carbon-aware optimizations in collaborative MR. In doing so, we make the following contributions. 
\begin{enumerate}[leftmargin=*]
    \item Outline the ecosystem of MR applications and analyze the major energy/carbon footprint sources in MR pipelines.
    \item Identify the opportunities for reducing the energy/carbon footprint and highlight the tradeoffs that must be navigated. 
    \item Outline research directions for researchers, application developers, and end users to enhance MR's sustainability. 
\end{enumerate}

\section{Landscape of Collaborative MR}
\label{sec:background}
\begin{figure}[t]
    \centering
    \includegraphics[width=0.96\linewidth]{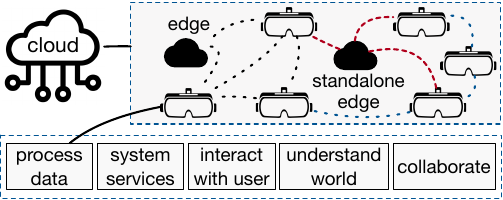} 
    \vspace{-0.3cm}
    \caption{\emph{\textbf{An overview of the collaborative mixed reality landscape.}}}
    \vspace{-0.6cm}
    \label{fig:overview}
\end{figure}

This section overviews the collaborative mixed reality (MR) components and sources of energy consumption and carbon emissions. As illustrated in Figure~\ref{fig:overview}, the MR ecosystem consists of MR devices connected to a local network, an edge, or a cloud in various configurations. Across these tiers, the MR sustainability implications include \emph{embodied carbon emissions} in the hardware supply chain and \emph{software operational emissions} due to electricity use. Next, we briefly describe its hardware and software components. 

\vspace{0.05cm}
\noindent
\textbf{\textsc{Hardware Components}}
include the headset and physical components in the network, at the edge, or in the cloud. 

\noindent
\textbf{1. {Headsets}} are a user's primary interface to the collaborative MR ecosystem. They include sensors, a display, processors, networking components, and a battery. Similar to mobile phones, embodied emissions dominate the lifecycle carbon footprint of MR headsets~\cite{Elgamal:2023:Design}, which end users and application developers cannot change. However, operational emissions are significant and will increase as advances in battery technology or wireless power transfer extend daily usage time. 

\noindent
\textbf{2. {Network infrastructure}} connects headsets to the edge or cloud servers, and its energy consumption depends on the data transfer requirements and the distance between endpoints. While the network's carbon emissions can be significant, prior work has not been done to quantify and reduce them. 

\noindent
\textbf{3. {Edge computing}} is pivotal to enabling real-time MR applications by providing high processing power of a dozen to hundreds of powerful servers close to the end user. While the sustainability implications of edge computing vary depending on the energy source, such as diesel generator-powered edge vs. solar-powered edge, it is primarily used to enable latency-critical applications. The use of edge infrastructure also reduces energy consumption in the network.

\noindent
\textbf{4. {Cloud}} servers often handle the most resource-intensive MR tasks, and their data processing and storage emissions can be substantial. Despite their significant power needs, the cloud's system-level efficiency is often higher than the edge but lower than the headsets that employ energy-efficient embedded processors, such as ARM-based. However, using the cloud may be inevitable for some applications as smaller processors cannot fit the bigger artificial intelligence (AI) models.

\vspace{0.05cm}
\noindent
\textbf{\textsc{Software Components}} in collaborative MR are numerous; we outline essential tasks and the related work in Table~\ref{tab:mr-software-related-work}.

\noindent
\textbf{1. Data collection and pre-processing} includes key tasks such as offline sensor calibration~\cite{jadid2019utilizing-sensor-calibration} and synchronization~\cite{ganeriwal2003timing-sensor-synchronization} and online data filtering~\cite{han2024secure-datafiltering} before it is used to capture user interactions and render MR experiences. The energy intensity of these tasks depends on the complexity of the environment, the types of sensors used, and application data requirements.

\noindent
\textbf{2. System services} in MR systems provide fundamental services that maintain the operational efficiency of the device, such as display brightness adjustment. Managing power-aware system states~\cite{liu2023metaverse} and optimizing idle states conserve energy~\cite{lyu2023metavradar}, enhance performance, and extend battery life.

\noindent\textbf{3. User interaction} management includes gesture recognition, object manipulation and rendering, and display control.

Gesture recognition technologies such as eye-gaze tracking~\cite{sibert2000evaluation-eye-gaze-objects, Li:2017:LowPowerGaze}, hand gesture recognition~\cite{luong2023controllersvshand}, and voice commands~\cite{hanifa2021review-voice-recogin} require continuous tracking of users to enable interactions with the virtual environment aligned with human behaviors and expectations. The rendering of 3D objects~\cite{massa2016deep2d-ed} enhances user interaction. However, enabling immersive experiences requires many tasks, such as direct manipulation of virtual objects~\cite{not-my-hands}, simulating real-world physics to enable realistic interactions, proper alignment and anchoring of virtual objects in the real world, and occlusion handling to ensure virtual objects are consistent with the physical world~\cite{alfakhori2022occlusion}. 

Display management tasks such as resolution management~\cite{debattista2018frameratevs-resoultion}, color calibration~\cite{dash2021much-dsiplay-oled-embedded}, adaptive brightness~\cite{shye2009into}, and foveated rendering~\cite{FOVxiao2016augmenting}  ensure that the visual output is optimized for device capabilities and tailored to the user's viewing comfort and environmental conditions. These settings are crucial for maintaining clarity, color fidelity, and overall visual comfort to prolong engagement in MR environments.

However, the energy aspects of these interactions are ignored to provide a safe, immersive, and secure MR experience. 

\noindent\textbf{4. World understanding} tasks refine the system's ability to interpret the surroundings. Depth estimation measures distances and relationships between objects, which helps place virtual items accurately in a real space~\cite{depth-perception-diaz2017designing}. Object segmentation~\cite{hao2020brief-semantic-segmnetation} and detection~\cite{object-size-chen2022effects} identify and categorize different environmental elements, allowing the system to interact intelligently. These capabilities make MR more useful for practical applications, blending digital content seamlessly with the physical world~\cite{barhorst2021blending} and enhancing user interaction by accurate head and pose tracking~\cite{wang2021tartanvo}. However, it is worth noting that most of the improvements for these services leverage compute-intensive deep learning approaches.

\noindent\textbf{5. Collaboration} is crucial in enabling immersive user experiences through effective content sharing across devices and with cloud/edge requiring services such as content delivery~\cite{zheng2023content-delivery} and content caching~\cite{zhang2018cooperative-content-cache}. It also requires managing computational loads efficiently by processing them at the edge~\cite{leng2019energy-video-processing-compression} or cloud~\cite{kim2022edge-cloud-based-processing}. The shared remote experiences~\cite{gonzalez2023virtual-remote-surgery} require blending co-located users with remote participants~\cite{schafer2022survey-collabration}. In shared experiences, collaboration spans the entire ecosystem and is highly energy-intensive, with a further energy consumption challenge across multiple locations.

\section{Sustainable Collaborative MR}
\label{sec:scope}
MR sustainability can be improved by reducing the energy and carbon footprint or exploiting potential tradeoffs between energy, carbon, and performance. We expand Table~\ref{tab:mr-software-related-work} to outline energy efficiency improvement (\texttt{EEI}), carbon efficiency improvement (\texttt{CEI}), energy performance tradeoff (\texttt{EPT}), carbon-energy tradeoff (\texttt{CET}), and carbon-performance tradeoff (\texttt{CPT}) opportunities in relevant tasks.

\vspace{0.1cm}
\noindent
\textbf{\textsc{Improving Efficiency}} refers to the opportunities for reducing energy consumption and carbon footprint of MR without impacting performance, potentially at higher cost. 

While the energy efficiency of computing has significantly improved, there are further optimizations possible in computing hardware's energy efficiency, software's algorithmic efficiency, and hardware-software co-design~\cite{Elgamal:2023:Design}. As outlined in Table~\ref{tab:mr-software-related-work}, improving energy efficiency may be difficult for the tasks requiring significant performance gains, as these tasks are likely to use more computationally intensive methods. Additionally, headset vendors must give application developers more control over the hardware to enable application-specific and context-aware optimizations.

The operational carbon footprint of MR depends on the carbon intensity of electricity used to power the headset, edge, and cloud. The carbon intensity depends on the mix of generation resources used to generate the electricity. If fossil-fuel-based power plants generate electricity, the carbon intensity would be high and show less variability. When there is no variability in the carbon intensity, carbon efficiency, and energy efficiency are the same. However, if carbon intensity shows spatiotemporal variability, the carbon and energy efficiencies diverge~\cite{Hanafy:2023:War}; it not only matters how much energy is consumed but also when and where it is consumed. Most tasks, except collaborative ones, have almost no flexibility and cannot explicitly optimize carbon-efficiency.

\vspace{0.07cm}
\noindent
\textbf{\textsc{Carbon, Energy, and Performance Tradeoffs}} must be navigated to improve MR sustainability if energy- and carbon-efficiency improvements have been exploited. 

MR's performance in creating immersive experiences for users takes the front seat, but not all applications require the highest level of resources, and significant sustainability gains can be made with a favorable performance sacrifice. Modern MR applications create seamless interactions and realistic simulations, requiring detailed and interactive virtual environments based on complex rendering tasks. However, the bigger is not always better, and immersiveness gains from high-resolution designs can be marginal for many applications. 
The application can afford better energy and performance tradeoffs (\texttt{EPT}) by prioritizing function over form.

Electricity's carbon efficiency exhibits spatiotemporal variations, and the energy efficiency of different components in the MR ecosystem varies.  These variations can be leveraged to gracefully navigate the carbon and energy tradeoffs (\texttt{CET}). For example, many offline tasks, such as compression and caching, can be offloaded to low-carbon edge/cloud locations where they run on low-carbon electricity. While the headset and the network consume energy when transferring the data, the total energy consumed can still be smaller than the on-device processing's overall carbon footprint. However, sustainability implications beyond energy and carbon must be considered, as water consumption and computing resource requirements can be significant. These tradeoffs are especially possible in collaborative MR scenarios, \emph{where users may already be geographically distributed and have carbon intensity variations.} 

Finally, the carbon-performance tradeoffs (\texttt{CPT}) can be exploited in multiple phases of the MR lifecycle. Prior work on design space optimization demonstrates that significant hardware-software co-design opportunities can help reduce the hardware requirement in MR headsets~\cite{Elgamal:2023:Design}. The reduced hardware specification significantly reduces the headsets' embodied carbon footprint while improving operational efficiency. Similar opportunities exist for optimizations across the ecosystem, such as headset vs edge. It can also be exploited by offline and background tasks with loose latency requirements, e.g., data compression or trend analytics. These tasks can wait for low-carbon periods to run, which will reduce carbon footprint but increase the completion time for these tasks.

We acknowledge that our list of potential research directions is not exhaustive, and other opportunities to improve MR's sustainability exist. However, our work takes an important step towards making MR sustainable and integrating sustainability as an optimization metric in MR research.

\def\arraystretch{1}
\setlength{\tabcolsep}{6pt}
\begin{table}[t]
\caption{Major tasks and related work in MR with potential for energy efficiency improvement (\texttt{EEI}), carbon efficiency improvement (\texttt{CEI}), energy performance tradeoff (\texttt{EPT}), carbon-energy tradeoff (\texttt{CET}), and carbon-performance tradeoff (\texttt{CPT}).
}
\vspace{-0.25cm}
\centering
\footnotesize
\centering
\begin{tabular}{|@{\hspace{0.1cm}}p{.4cm}|@{\hspace{0.05cm}}p{4.7cm}|@{\hspace{0.1cm}}p{.3cm}|@{\hspace{0.1cm}}p{.3cm}|@{\hspace{0.1cm}}p{.3cm}|@{\hspace{0.1cm}}p{.3cm}|@{\hspace{0.1cm}}p{.3cm}|}\hline

& 
\textbf{Tasks} & 
\textbf{\texttt{EEI}} &
\textbf{\texttt{CEI}} &
\textbf{\texttt{EPT}} & 
\textbf{\texttt{CET}} & 
\textbf{\texttt{CPT}} \\ \hline \hline 

\multicolumn{7}{|l|}{\textbf{\cellcolor{gray!20} 1 -- Data collection and processing}}\\ \hline 

\textbf{1.1} & Sensor calibration \cite{jadid2019utilizing-sensor-calibration} & $\checkmark$& -- &$\checkmark$ &-- & -- \\ \hline

\textbf{1.2} & Sensor synchronization \cite{ganeriwal2003timing-sensor-synchronization}  & $\checkmark$& -- &$\checkmark$ &-- & --\\ \hline

\textbf{1.3} & Data filtering~\cite{han2024secure-datafiltering} & $\checkmark$& -- &$\checkmark$ &-- & --\\ \hline \hline 

\multicolumn{7}{|l|}{\textbf{\cellcolor{gray!20}2 -- System services}}\\ \hline

\textbf{2.1} & Power-aware system states~\cite{liu2023metaverse}                         & $\checkmark$ & $\checkmark$  &  $\checkmark$ & -- & --  \\ \hline

\textbf{2.2} & Idle state optimization~\cite{lyu2023metavradar}                           & $\checkmark$ & $\checkmark$ & $\checkmark$ & -- & --  \\ \hline \hline

\multicolumn{7}{|l|}{\textbf{\cellcolor{gray!20}3 -- User interaction}}\\ \hline 

 \multicolumn{7}{|l|}{\hspace{0.55cm}\textbf{Gesture recognition}}\\ \hline

\textbf{3.1} & Eye-gaze tracking \cite{Li:2017:LowPowerGaze} & $\checkmark$& -- & -- &-- & -- \\ \hline

\textbf{3.2} & Hand gesture recognition \cite{luong2023controllersvshand} & -- &-- & -- & -- &--   \\ \hline

\textbf{3.3} & Voice recognition \cite{hanifa2021review-voice-recogin} & -- &-- & -- & -- &-- \\ \hline

 \multicolumn{7}{|l|}{\hspace{0.55cm}\textbf{Object rendering and manipulation}}\\ \hline

\textbf{3.4} & Rendering 2D/3D models~\cite{massa2016deep2d-ed} & --&-- & -- & -- &--  \\ \hline

\textbf{3.6} & Direct object manipulation \cite{not-my-hands} & -- &-- & -- & -- &-- \\ \hline

\textbf{3.5} & Anchoring, aligning, \& persistence~\cite{alfakhori2022occlusion} &  --&-- & -- & -- &-- \\ \hline

 \multicolumn{7}{|l|}{\hspace{0.55cm}\textbf{Display}}\\ \hline
 
\textbf{3.7} & Resolution management \cite{debattista2018frameratevs-resoultion} & $\checkmark$& -- &$\checkmark$ &-- & --  \\ \hline

\textbf{3.8} & Color calibration \cite{dash2021much-dsiplay-oled-embedded} & $\checkmark$& -- &$\checkmark$ &-- & --   \\ \hline

\textbf{3.9} & Adaptive brightness \cite{shye2009into} & $\checkmark$& -- &$\checkmark$ &-- & -- \\ \hline
\textbf{3.10} & Foveated rendering \cite{FOVxiao2016augmenting} & $\checkmark$& -- &$\checkmark$ &-- & -- \\ \hline \hline

\multicolumn{7}{|l|}{\textbf{\cellcolor{gray!20}4 -- Understanding the world}}\\ \hline 

 \multicolumn{7}{|l|}{\hspace{0.55cm}\textbf{Scene understanding}}\\ \hline 

\textbf{4.4} & Depth estimation~\cite{depth-perception-diaz2017designing} & $\checkmark$& -- & -- &-- & -- \\ \hline

\textbf{4.6} & Semantic segmentation~\cite{hao2020brief-semantic-segmnetation} &  -- &-- & -- & -- &-- \\ \hline 

\textbf{4.5} & Object detection~\cite{object-size-chen2022effects} & $\checkmark$& $\checkmark$ & -- &-- & -- \\ \hline

\multicolumn{7}{|l|}{\hspace{0.55cm}\textbf{Spatial mapping and 3D reconstruction}}\\ \hline

\textbf{4.1} & Handle occlusion, avoid collision~\cite{alfakhori2022occlusion}& -- &-- & -- & -- &-- \\ \hline

\textbf{4.2} & Real \& virtual world blending~\cite{barhorst2021blending}  & -- &-- & -- & -- & \\ \hline

\textbf{4.3} & Pose/head tracking~\cite{wang2021tartanvo}& -- &-- & -- & -- &-- \\ \hline \hline

\multicolumn{7}{|l|}{\textbf{\cellcolor{gray!20}5 -- Collaboration}}\\ \hline

 \multicolumn{7}{|l|}{\hspace{0.55cm}\textbf{Network/edge offloading}}\\ \hline

\textbf{5.1} & Content delivery~\cite{zheng2023content-delivery} & $\checkmark$& $\checkmark$ &$\checkmark$ &-- & -- \\ \hline

\textbf{5.2} & Content caching on edge ~\cite{zhang2018cooperative-content-cache}& $\checkmark$& $\checkmark$ &$\checkmark$ &-- & $\checkmark$  \\ \hline

\textbf{5.3} & Compression~\cite{leng2019energy-video-processing-compression} & $\checkmark$& $\checkmark$ &$\checkmark$ &-- & $\checkmark$  \\ \hline

\textbf{5.4} & Cloud-based processing~\cite{kim2022edge-cloud-based-processing}& $\checkmark$& $\checkmark$ &$\checkmark$ &$\checkmark$ & $\checkmark$ \\ \hline

\multicolumn{7}{|l|}{\hspace{0.55cm}\textbf{Multi-user experience}}\\ \hline

\textbf{5.5} & Remote experiences~\cite{gonzalez2023virtual-remote-surgery} & $\checkmark$ &-- & -- & $\checkmark$ & $\checkmark$ \\ \hline

\textbf{5.6} & Blending co-located \& remote users~\cite{schafer2022survey-collabration} & $\checkmark$ &-- & -- & -- &-- \\ \hline \hline
\end{tabular}
\vspace{-0.7cm}
\label{tab:mr-software-related-work}
\end{table} 

\section{Conclusion and Future Work}
\vspace{-0.2cm}
\label{sec:cons}
MR's environmental sustainability implications are growing as it is deployed for applications beyond leisure. Prior work has explored specific aspects of MR sustainability, but significant opportunities remain, especially in the collaborative MR ecosystem. We map the collaborative MR landscape and discuss potential opportunities and their implications. We posit that efficient MR needs to integrate lessons from broader computing with targeted innovations tailored to the unique demands of MR systems. While there have been improvements, future work should optimize hardware to support novel software algorithms, enhancing sustainability and user experience. Each step forward contributes to more energy-efficient MR technologies and aligns with broader goals to reduce the carbon footprint of digital systems globally.

\vspace{-0.25cm}
\section*{Acknowledgements}
\vspace{-0.25cm}
This research is supported by NSF Grants 2105564, 2236987, 2346133, 2237485, 2230143, and VMware.


\end{document}